\documentclass[12pt]{iopart}

\usepackage{amsfonts}
\usepackage{float}
\usepackage{subfigure}
\usepackage[pdf]{pstricks}
\usepackage{graphicx}
\usepackage{epstopdf}
\usepackage{epsfig}
\usepackage{verbatim}
\usepackage{cite}

\newcommand{\eq}[1]{\begin{equation}#1\end{equation}}
\newcommand{\eqa}[1]{\begin{eqnarray}#1\end{eqnarray}}

\begin{document}

\title{Two-mode squeezed states in the $q$-deformed Pegg-Barnett Fock space }

\author{Yidan Wang}

\address{School of Physical \& Mathematical Sciences, 
Nanyang Technological University, \\
SPMS-04-01, 21 Nanyang Link, 
Singapore 637371}
\ead{yidanhaha@gmail.com} 

\author{Leong Chuan Kwek}
\address{Centre for Quantum Technologies, National University of Singapore, 2 Science
Drive 3, Singapore 117542 \\ National Institute of Education and Institute of Advanced Studies, Nanyang
Technological University, 1 Nanyang Walk, Singapore 637616}
\ead{kwekleongchuan@nus.edu.sg}

\begin{abstract}
We study the coherent state and two-mode squeezed state in the $q$-deformed Pegg-Barnett(PB) formalism. We show that when the truncation of the Fock space $S$ is large enough, the phase properties of the $q$-deformed PB coherent state approach that of the undeformed PB coherent state. We also investigate the entanglement properties of the two-mode squeezed states in both the $q$-deformed and undeformed PB Fock space with the real squeezing parameter $r$.  We see that if $S$ is sufficiently large, the conventional two-mode squeezed states can be approximated with the PB (deformed and undeformed) states for arbitrary $r$. However, the value of $S$ required increases more rapidly with the $q$-deformed PB states than the PB states as a function of $r$.
\end{abstract}


\maketitle

\section{Introduction}
The possible existence of a Hermitian phase operator for the harmonic oscillator (or a single mode of the electromagnetic field) has intrigued physicists  since the start of quantum mechanics. 
A natural assumption based on the correspondence between the classical Poisson bracket and quantum commutation relations leads to a canonical commutation relation for the optical phase operator $\phi$ and number operator:
\begin{equation}
[\phi, N]=-i, 
\end{equation}
first suggested by Dirac \cite{Dirac27}.
However, the matrix elements of $\phi$ in a number state basis are undefined:
\eq{
(n-n')\langle n'|\phi| n\rangle =-i \delta_{nn'}
}
It was shown by Louisell \cite{Louisell63} and Susskind and Glogower \cite{Susskind64} that the number-phase commutator is not consistent with the existence of a well-defined Hermitian phase operator. Susskind and Glogower \cite{Susskind64} also realized that the semi-boundedness of the Fock space, i.e., the existence of a cut-off at the vacuum state $|0\rangle$ of the spectrum of the number operator, is the main reason for the difficulty in the definition of a Hermitian phase operator. One possible solution to the problem, as proposed by Pegg and Barnett \cite{Barnett86, Barnett89}, is to define the Hermitian phase operator in a finite-dimensional Fock space. They have shown that the phase operator can be rigorously constructed in the finite-dimensional space and the infinite Fock space can be regarded as an infinite limit of the finite space. However, the Pegg-Barnett formalism is not a 
closed associative oscillator algebra. \\

The $q$-deformed harmonic oscillator can also give a finite Fock  
space when $q=exp(\frac{2\pi i}{S+1})$ for a positive integer $S$. In Ref.\cite{Ellinas91},  
a Hermitian phase operator based on $q$-deformed finite Fock space was defined 
in a similar way to Pegg and Barnett. The phase operator in the conventional infinite Fock space is obtained by taking $S \rightarrow \infty$. In this limit, the q-deformation reverts to the usual Heisenberg-Weyl Lie algebra.
In our paper, we wish to construct the coherent state and two-mode squeezed state in the context of $q$-deformed Hermitian phase space construction and explore the entanglement properties of the two-mode squeezed states. We would also like to test whether the coherent state and the $q$-deformed two-mode squeezed state will approach the conventional states when one takes the limit $S\rightarrow \infty$ ($q\rightarrow 1$). 


The paper is organized as follows. A brief review of the Susskind-Glogower (SG) phase operator and Pegg-Barnett (PB) Hermitian phase operator is given in \sref{PB}. The $q$-deformed algebra and the construction of phase operator in $q$-deformed PB Fock space is given in \sref{qHP}. In \sref{CoPB}, the coherent state in the $q$-deformed PB Fock space is constructed and its phase properties will be studied.  In \sref{SqPB}, the two-mode squeezed state in the $q$-deformed PB Fock space is constructed and its entanglement entropy is compared to the two-mode squeezed state in the PB Fock space.

\section{Pegg-Barnett formalism of Hermitian phase operator}\label{PB}
Starting with the classical electromagnetic wave amplitude $A=N^{1/2} \exp(i\phi)$ where $N^{1/2}$ is the modulus of the amplitude and $\phi$ is the phase, Susskind and Glogower defined the phase operator in terms of the polar decomposition of $a$ and $a^{\dagger}$:  \cite{Susskind64}
\numparts
\eqa{
a&= &\exp(i\phi)N^{1/2},\\
a^{\dagger}&= &N^{1/2}\exp(-i\phi).
}
\endnumparts
The exponential phase operator of Susskind and Glogower is 
\numparts
\eqa{
\exp(i\phi)&=&\sum_{n=0}^{\infty}|n\rangle\langle n+1|,\\
\exp(-i\phi)&=&\sum_{n=0}^{\infty}|n+1\rangle \langle n|.
}
\endnumparts
$\exp(i\phi)$ and $\exp(-i\phi)$ do not commute and are not unitary. Therefore the Susskind-Glogower formalism does not give a Hermitian phase operator $\phi$ \cite{Barnett86}. We will call this infinite Fock space \emph{SG Fock space} and the states defined on it SG states.

Pegg and Barnett attacked the problem of Hermitian phase operator from a different approach.  They noticed that the eigenstates of the phase operator  can be defined as a superposition of the Fock (number) states when the Fock space is finite \cite{Barnett89}.  In the paper, we call the finite Fock space $\{|n\rangle^S\}$ \emph{PB Fock space}. $|S\rangle^S$ is the highest number state and $S$ is a positive integer number. The annihilation operator $\tilde{a}$ and creation operator $\tilde{a}^\dagger$ act on $\{|n\rangle^S\}$ the same way as $a$ and $a^\dagger$ do on the SG Fock space except that $\tilde{a}^\dagger$ acts on $|S\rangle^S$ gives 0.
\numparts
\eqa{
\tilde{a}^S|n\rangle^S &=&\sqrt{n}|n-1\rangle^S \ \mbox{for} \ 1\leq n\leq S, \qquad \tilde{a}|0\rangle^S = 0,\\
\tilde{a}^\dagger |n\rangle^S &=&\sqrt{n+1}|n+1\rangle^S\ \mbox{for}\ 0\leq n\leq S-1,
 \qquad \tilde{a}^\dagger|S\rangle^S = 0.
}
\endnumparts
The number operator in this PB Fock space is
\eq{
\tilde{N}= \sum_{n=1}^S n|n\rangle^S {}^S\langle n|,
} 
Pegg and Barnett defined the phase eigenstate $|{\theta}\rangle^S$ to be \cite{Barnett89}
\eq{
|{\theta}\rangle^S=\sqrt{\frac{1}{S+1}}\sum_{n=0}^{S}
\exp(in\theta)|n\rangle^S.
}
where $\theta$ can take any value in the range $\theta_0$ to $\theta_0 + 2\pi$, where $\theta_0$ is a reference phase. There are an uncountably infinite different phase states and they form an overcomplete basis for $\{|n\rangle^S\}$ \cite{Barnett89}. However, a complete orthonormal basis can be chosen by selecting a subset of these states. Such a subset will consist of a reference phase state
\eq{
| \theta_0\rangle^S=\sqrt{\frac{1}{S+1}} \sum_{n=0}^{S}\exp(in\theta_0)|n\rangle^S
}
together with the $S$ phase states that are orthogonal to it (and to each other). The basis states are then $|\theta_m\rangle$, where
\eq{
\theta_m=\theta_0+\frac{2\pi m}{S+1},\ m=0,1,\ldots ,S.
}
The value of $\theta_0$ is arbitrary and it determines the particular phase state basis up to 
$\theta_0\ mod\ 2\pi$.
The number state $|n\rangle^S$ can be expanded in terms of the phase state basis $|\theta_m\rangle^S$ as
\eqa{\nonumber
|n\rangle^S&=&\sum_{m=0}^{S}|\theta_m\rangle^S{}^S\langle\theta_m|n\rangle^S\\
&=&\sqrt{\frac{1}{S+1}}\sum_{m=0}^{S}\exp(-in\theta_m)|\theta_m\rangle^S.
}
The Hermitian phase operator $\tilde{\phi}$ is required to have eigenvectors $|\theta_m\rangle^S$ and corresponding eigenvalues $\theta_m$. It can be constructed as 
\eqa{\nonumber
\tilde{\phi}&=&\sum_{m=0}^{S}\theta_m|\theta_m\rangle^S{}^S\langle\theta_m|\\
&=&\theta_0+\sum_{m=0}^{S}\frac{2\pi m}{S+1}|\theta_m\rangle^S{}^S\langle\theta_m|
}
The phase operator depends on the choice of reference phase $\theta_0$ and its eigenvalues range from $\theta_0$ to $\theta_0+2\pi$.\\
The Fock state representation of $\tilde{a}$ and $\tilde{a}^\dagger$ is given by
\numparts
\eqa{
\tilde{a}&=&\displaystyle \sum_{n=1}^{S} \sqrt{n}|n-1\rangle^S {}^S\langle n|, \\
\tilde{a}^\dagger &=& \sum_{n=1}^{S} \sqrt{n}|n\rangle^S {}^S\langle n-1|, \label{eqa}
}
\endnumparts
and we can get the commutation relation of  $\tilde{a}$ and $\tilde{a}^\dagger$:
\eq{
[\tilde{a},\tilde{a}^\dagger]=1-S|S\rangle^S {}^S\langle S|
}
The Pegg-Barnett formalism do not close into an associative algebra since $[\tilde{a},\tilde{a}^\dagger]$ does not commute with the Hamiltonian $\tilde{N}=(\tilde{N}+\frac{1}{2})\hbar \omega$ \cite{Arik99}.\\
The Pegg-Barnett construction also gives a polar decomposition of $\tilde{a}$ and $\tilde{a}^{\dagger}$ \cite{Barnett89}
\numparts
\eqa{
\tilde{a} & = & \exp(i \tilde{\phi})\tilde{N}^{1/2}\\
\tilde{a}^{\dagger} & = & \tilde{N}^{1/2}\exp(-i{\tilde{\phi}})
}
\endnumparts
The phase eigenstates and phase operators in the SG Fock space can be seen as the limit of $|\theta\rangle^S$ and $\tilde{\phi}$ when $S\rightarrow \infty$.

\section{The $q$-deformed Hermitian phase operator}\label{qHP}
In the Pegg-Barnett formalism, a finite-sized Fock space is crucial for the definition of the eigenstates of the Hermitian phase operator.  As we will show below, the $q$-deformed oscillator algebra with  a special choice of $q$ ( $q=$exp$(\frac{2\pi i}{S+1})$, $S \in \mathbb{Z}$ ) also gives a finite Fock space, thus a Hermitian phase operator can be similarly constructed on it, first shown by Ellinas \cite{Ellinas91}. The phase operator in the SG Fock space can be obtained by taking the limit $S\rightarrow \infty$. The theoretical advantage of constructing phase operator using $q$-deformed algebra is that it closes the oscillator algebra. 

For the $q$-deformed harmonic oscillator, the commutation relations among the $q$-deformed creation and annihilation operators $a_q$, $a^{\dagger}_q$ and the number operator $N_q$ are \cite{Nelson94,Biedenharn89,Harouni09}
\numparts
\eqa{
a_qa_q^{\dagger}-q a_q^{\dagger} a_q&=&q^{-N_q}\\ 
\left[N_q,a_q \right] &=&  -a_q \\ 
\left[N_q,a^\dagger_q \right] &=& a^\dagger.
}
\endnumparts
Here $a_q^{\dagger}$ and $a_q$ raises and lowers the $q$-deformed number states $|n\rangle_q$ in a similar way as $a$ and $a^\dagger$ do in the SG Fock space, except for the ``distortion" in the coefficient:
\eqa{
a_q^{\dagger}|n\rangle_q=\sqrt{[n+1]}|n+1\rangle_q, \ \ a_q|n\rangle_q=\sqrt{[n]}|n\rangle_q. 
}
where $[n]=\frac{q^{n}-q^{-n}}{q-q^{-1}}$.\\
\noindent Since $[0]=0$, $a_q$ acting on the ground state $|0\rangle_q$ will return vacuum. 
\eqa{
a_q|0\rangle_q=0.
}
The $q$-deformed number operator $N_q$ and creation and annihilation operators $a_q$ and $a^{\dagger}_q$ can be defined as
\numparts
\eqa{
N_q&=&\sum_{n=0}^{\infty} n|n\rangle_q{}_q\langle n|,\\ 
a_q&=&\displaystyle \sum_{n=0}^{\infty} \sqrt{[n]}|n-1\rangle_q {}_q\langle n|, \\ 
a_q^\dagger &=& \sum_{n=0}^{\infty} \sqrt{[n]}|n\rangle_q {}_q\langle n-1|, \label{eqa}
}
\endnumparts
Note that when $q=$exp$(\frac{2\pi i k}{S+1})$, where $k$ and $S+1$ are coprime integers and $S$ is non-negative, $[S]=0$. This means $a^\dagger_q$ acting on the state $|S\rangle_q$ will return vacuum and therefore $|S\rangle_q$ is the highest state of the $q$-deformed Fock space. We can therefore construct the Hermitian phase operator upon this finite $q$-deformed Fock space. For simplicity, we choose $k=1$ and $q=$exp$(\frac{2\pi i}{S+1})$. This finite $q$-deformed Fock space is called  \emph{ $q$-deformed PB Fock space} and is denoted as $\{|n\rangle_q^S\}$.\\

The number operator $\tilde{N}_q$ and the creation and annihilation operators $\tilde{a}_q$ and $\tilde{a}^{\dagger}_q$ on the $q$-deformed PB Fock space can be represented as 
\numparts
\eqa{
\tilde{N}_q&=&\sum_{n=0}^{S} n|n\rangle_q^S{}^S_q\langle n|,\\ 
\tilde{a}_q&=&\displaystyle \sum_{n=1}^{S} \sqrt{[n]}|n-1\rangle_q^S {}^S_q\langle n|, \\ 
\tilde{a}_q^\dagger &=& \sum_{n=1}^{S} \sqrt{[n]}|n\rangle_q^S {}^S_q\langle n-1|, \label{eqa}
}
\endnumparts

\noindent The definition of the phase eigenstates upon $\{|n\rangle_q^S\}$ is very similar to that in PB Fock space:
\eqa{
|\theta\rangle_q^S &=&\sqrt{\frac{1}{S+1}}\sum_{n=0}^{s}
\exp (in\theta)|n\rangle_q^S.
}
The $q$-deformed phase operator $\tilde{\phi}_q$ with reference eigenstate $|\theta_0\rangle_q^S$ is similarly defined as
\eqa{
\nonumber
\tilde{\phi}_q&=&\sum_{m=0}^{S}\theta_m|\theta_m\rangle_q^S{}^S_q\langle\theta_m|\\
&=&\theta_0+\sum_{m=0}^{S}\frac{2\pi m}{S+1}|\theta_m\rangle_q^S{}^S_q\langle\theta_m|
}
with $\theta_m=\theta_0+\frac{2\pi m}{S+1}$.
The polar decomposition of $\tilde{a}_q$ and $\tilde{a}_q^\dagger$ in terms of the unitary phase operator and the number operator is
\numparts
\eqa{
\tilde{a}_q&=&\exp(i\tilde{\phi}_q)\sqrt{[\tilde{N}_q]}\\
\tilde{a}_q^{\dagger}&=&\sqrt{[\tilde{N}_q]}\exp(-i{\tilde{\phi}_q})
}
\endnumparts

\section{The $q$-deformed coherent state in PB\label{CoPB}}
The coherent state on the SG Fock space is a state that is closest to a classical state, and it is a minimum uncertainty state with equal variances in both quadratures. 
It is also created by the action of the displacement operator on the vacuum state
\eq{
|\alpha\rangle = e^{\alpha a^{\dagger}-\alpha^* a}|0\rangle, \label{cohe1}
}
where $\alpha\in \mathbb{C}$,
and it is the eigenstate of the annihilation operator $a$:
\eq{
a|\alpha\rangle =\alpha |\alpha\rangle. \label{cohe2}
}
Now we will discuss the coherent state in PB Fock space, in $q$-deformed Fock space, and in $q$-deformed PB Fock space.

In the PB Fock space $\{|n\rangle^S\}$, the coherent state $|\alpha\rangle^S$ can be constructed by truncating the terms higher than $|S\rangle^S$ \cite{Zhu94}. 
\eqa{
|\alpha\rangle^S &=& A \sum_{n=0}^{S}\frac{\alpha^n}{\sqrt{n!}}|n\rangle^S. \\ \label{cohe2}
}
In the $q$-deformed Fock space $\{|n\rangle_q\}$, we can define the coherent state $|\alpha\rangle_q$ to be the eigenvector of $a_q$ \cite{Nelson94}:
\eq{
a_q|\alpha\rangle_q =\alpha |\alpha\rangle_q. \label{coheq}
}
and it can be represented as
\eq{
|\alpha\rangle _q= A \sum_{n=0}^{\infty} \frac{\alpha ^n}{\sqrt{[n]!}} |n\rangle _q \label{cohe3}
}
where $A$ is the normalization constant.\\
In $q$-deformed PB Fock space, $q=\exp(\frac{2\pi i}{S+1})$. We can construct the coherent state by identifying the set of states $|n+k(S+1)\rangle_q$ ($k\in \mathbb{Z}$) in $\{|n\rangle^S\}$ with $|n\rangle_q^S$ in $\{|n\rangle_q^S\}$ and assign the sum of the coefficients of $|n+k(S+1)\rangle_q$ to be the coefficient of $|n\rangle_q^S$.\\
Let
\eqa{
|\alpha\rangle_q^S &=&\sum_{n=0}^{S}A(n)|n\rangle_q,
}
then
\eqa{
A(n)&=&A'\sum_{k=0}^{\infty} \nonumber
A\frac{\alpha^{ \left(n+k(S+1)\right)}}{\sqrt{[n+k(S+1)]!}}\\
&=&A'\left(\sum_{k=0}^{\infty} 
\frac{\alpha^{ \left(k(S+1)\right)}}{\sqrt{[k(S+1)]!}} 
\right)\frac{\alpha^n}{\sqrt{[n]!}}
}
where $A'$ is the normalization factor. Since $\sum_{k=0}^{\infty} 
\frac{\alpha^{ \left(k(S+1)\right)}}{\sqrt{[k(S+1)]!}}$ does not depend on $n$, the $q$-deformed PB coherent state can be simplified to
\eqa{
|\alpha\rangle^S_q &=& A''\sum_{n=0}^{S}\frac{\alpha^n}{\sqrt{[n]!}}|n\rangle_q^S. \label{coheqPB}
}
where $A''$ is the normalization constant. 
An alternative way to define $|\alpha\rangle_q^S$  is to truncate the summation in \eref{cohe3} at $|S\rangle_q$ where $[S]=0$ and $\frac{\alpha^S}{\sqrt{[S]!}}$ is not well-defined.
These two methods will give the same expression of $|\alpha\rangle_q^S$ in \eref{coheqPB}. \\
\subsection{Phase property
 of $q$-deformed PB coherent state}
In order to check whether the $q$-deformed PB formalism approaches the PB formalism for $S\rightarrow \infty$, we investigate whether the phase probability distribution of the coherent states defined in the $q$-deformed Fock space approaches their SG  counterparts when $S\rightarrow \infty$. To study the phase property of $|\alpha\rangle_q^S$ as $S$ increases, we plot the modulus of its decomposition coefficients $C_m$ in the phase eigenstates $|\theta_m\rangle^S_q$:
\eqa{
|\alpha\rangle_q^S&=&C_m |\theta_m\rangle^S_q. 
} as a function of $\theta_m$. 
This gives a distribution in the angle range $[-\pi,\pi)$.
The distribution depends on the choice of $\theta_0$. However,  different choices of $\theta_0$ only differ in a translation of the distribution.

\begin{figure}[H]
\begin{center}
\includegraphics[scale=0.7]{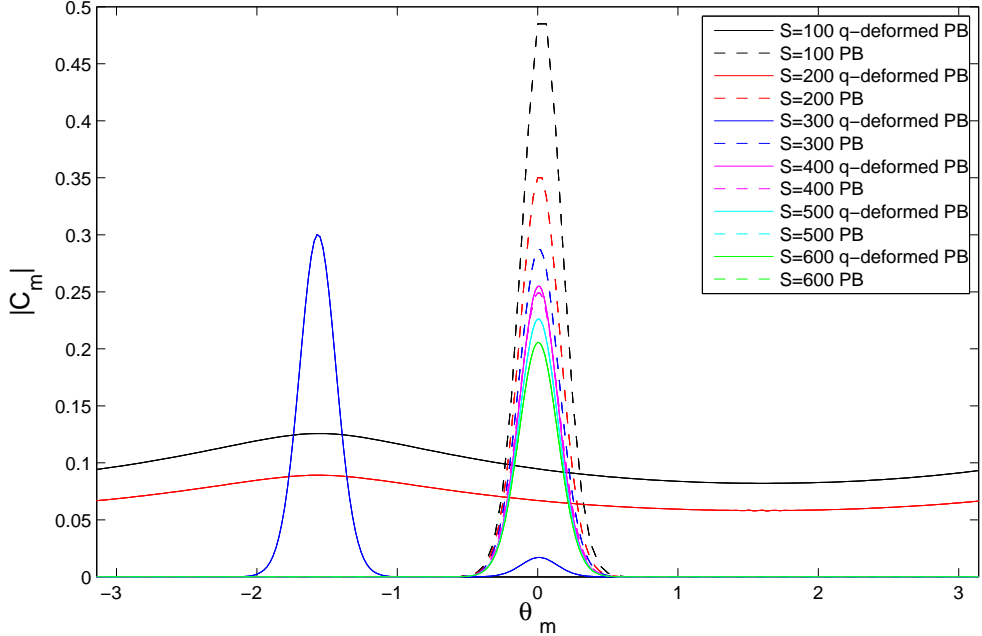}\caption{\label{fig:phase}
Coefficients $C_m$ of $|\alpha\rangle_q^S$ and $|\alpha\rangle^S$ in their respective phase eigenstate basis $\{|\theta_m\rangle_q^S\}$ and $\{|\theta_m\rangle^S\}$ with $\theta_0=-\pi$, $\alpha=5$.}
\end{center}
\end{figure}

In \fref{fig:phase}, the coefficients $|C_m|$ for $|\alpha\rangle_q^S$ are plotted for various choices of $S$ for $\alpha=5 $ and $\theta_0=-\pi$.
The coefficients of the PB coherent states $|\alpha\rangle^S$ in the phase eigenstate basis $\{|\theta_m\rangle^S\}$ are also plotted in dashed lines for a comparison. The choice of $\alpha$ and $\theta_0$ is the same as $|\alpha\rangle_q^S$.
\Fref{fig:phase} shows the approximation of $|\alpha\rangle_q^S$ for $|\alpha\rangle^S$ becomes increasingly better with increasing $S$, until the two curves eventually coincide for $S\geq500$. This suggests that for sufficiently large $S$, $|\alpha\rangle^S_q$ approaches $|\alpha\rangle^S$.

\section{$q$-deformed two-mode squeezed state in PB \label{SqPB}}
The two-mode squeezed state with a complex squeezing parameter $\zeta=re^{\theta}$ in SG is defined as
\begin{eqnarray}
\nonumber
|\zeta\rangle & = & e^{\zeta^{*}a_{1}a_{2}-\zeta a_{1}^{\dagger}a_{2}^{\dagger}}|0,0\rangle\\
 & = & \frac{1}{\cosh(r)}\sum_{n=0}^{\infty}\left(e^{i\theta}\tanh(r)\right)^n|n,n\rangle  
\end{eqnarray}
For a real squeezing parameter $r$,
\begin{eqnarray}
|r\rangle= \frac{1}{\cosh(r)}\sum_{n=0}^{\infty} \tanh^n(r)|n,n\rangle  \label{equcon}
\end{eqnarray}
Now we go on to discuss on the two-mode squeezed state defined on PB Fock space, $q$-deformed Fock space and $q$-deformed PB Fock space.
In the $q$-deformed PB formalism, the two-mode squeezed state can be constructed by truncating the terms higher than $|S,S\rangle^S$ 
\eq{
|r\rangle^S
= D\sum_{n=0}^{S}\tanh^n(r)|n,n\rangle^S, \label{r1}
}
where the normalization constant $D$ satisfy
\eq{
|D|^2=\left(\sum_{n=0}^{S}\tanh^{2n}(r)\right)^{-1}=\frac{1-\tanh^2(r)}{1-\tanh^{2S+2}(r)} 
\nonumber
}

In Ref.\cite{Meng08}, the authors have proposed a $q$-analogue
of the two-mode squeezed states with  real squeezing parameter $r$: \footnote{
An alternative definition of $q$-deformed squeezed state in Ref.\cite{Meng08} is
$$
|r\rangle_q' 
= C_{1}\sum_{n=0}^{\infty}\frac{n!\tanh^{n}(r)}{[n]!}|n,n\rangle_q, $$
However, the coefficients blow up very fast with increasing $n$ and the entanglement does not tend to the SG two-mode squeezed state when $q\rightarrow 1$.
}
\eq{
|r\rangle_q
= C\sum_{n=0}^{\infty}\frac{[n]!\tanh^{n}(r)}{n!}|n,n\rangle_q. \label{qsqstate}
}
where the normalization constant $C$ is given by
\eq{
|C|^{2}  =  \left(\sum_{n=0}^{\infty}\left(\frac{[n]!\tanh^{n}(r)}{n!}\right) ^2 \right)^{-1} 
\nonumber
}
When the $q$-deformed Fock space becomes the $q$-deformed PB Fock space, the summation  in  \eref{qsqstate} is no longer valid and we need to redefine the $q$-deformed two-mode squeezed state:
\eq{
|r\rangle_q^S =\sum_{n=0}^{S} C(n)|n,n\rangle_q^S.
}
Similar to the construction of $q$-deformed PB coherent states, there are also two schemes we can follow. We can either truncate the summation in \eref{qsqstate} at $n=S$ or identify the states $|n+(S+1)k, n+(S+1)k\rangle_q$ in the $q$-deformed Fock space with $|n,n\rangle_q^S$ in the $q$-deformed PB Fock space. 
Following the second scheme, $C(n)$ is obtained by adding the coefficients of $|n+(S+1)k,n+(S+1)k\rangle_q$ for different $k$.
\eqa{ \nonumber
C(n)&=&C'\sum_{k=0}^{\infty}\tanh(r)^{n+k(S+1)}\frac{[n+k(S+1)]!}{\left(n+k(S+1)\right)!}
\nonumber\\
&=&C'[n]!\sum_{k=0}^{\infty}\frac{\tanh(r)^{(n+(S+1)k)}([S+1]!)^k}{\left(n+k(S+1)\right)!}
}
where $C'$ is the normalization coefficient. Here we assume $[0]!=0$ and $[S+1]!=[S]!$.
Let $b=\tanh(r)\left([S+1]!\right)^{\frac{1}{S+1}}$, $C(n)$  becomes
\eqa{
C(n)&=&C'[n]!([S+1]!)^{\frac{-n}{S+1}}\sum_{k=0}^{\infty}\frac{b^{(n+(S+1)k)}}{\left(n+k(S+1)\right)!}
}
Now we go on to simplify 
\eqa{
K(n)&=&\sum_{k=0}^{\infty}\frac{b^{(n+k(S+1))}}{\left(n+k(S+1)\right)!}.
}
Since $(q^l)^n=e^{\frac{2\pi il}{S+1}n}$ is a periodic function of $n$ with period $S+1$, $e^{bq^l}$ can be expressed as
\eqa{
e^{bq^l} &=&  \sum_{n=0}^{\infty}\frac{b^n}{n!}q^{ln}\\
&=& \sum_{n'=0}^{S} \left(\sum_{k=0}^{\infty}\frac{b^{n'+k(S+1)}}{(n'+k(S+1))!}\right)q^{ln'}.
}
Note that the coefficients before $q^{ln'}$ is the expression $K(n')$ we want to evaluate.
By a clever superposition of $e^{bq^l}$ for different $l$, we can cancel the terms $q^{ln'}$ except for the $n'=n$ we want:
\eq{
q^{-ln}e^{bq^l}=\sum_{n'=0}^{S}
K(n')q^{l(n'-n)}
}
\eq{
\sum_{l=0}^{S}q^{-ln}e^{bq^l}=\sum_{n'=0}^{S}
K(n')\sum_{l=0}^Sq^{l(n'-n)}
\label{equaE}
}
Since
\eq{
\sum_{l=0}^S q^{l(n'-n)}=(S+1)\delta_{n',n},
}
\eref{equaE} becomes
\eq{
\sum_{l=0}^{S}q^{-ln}e^{bq^l}=
(S+1)K(n).
}
Therefore we have simplified the infinite series summation $K(n)$ to a finite series summation \footnote{
Following the first scheme of construction, the $q$-deformed PB Fock space is constructed by truncating   the summation in \eref{qsqstate} at $n=S$:
$$
|r^\prime \rangle_q^{S}= D_{1}\sum_{n=0}^{S}\frac{[n]!\tanh^{n}(r)}{n!}|n,n\rangle_q^S.
$$
Since the coefficient $\frac{[n]!\tanh^{n}(r)}{n!}$ drops very fast with increasing $n$, $|r^\prime \rangle_q^{S}$ is extremely close to $|r\rangle_q^S$.
}
:
\eq{
K(n)=\frac{1}{S+1}\sum_{l=0}^{S}q^{-ln}e^{bq^l}
}
\eq{
C(n)
=C'[n]!([S+1]!)^{\frac{-n}{S+1}}\frac{1}{S+1}\sum_{l=0}^{S}q^{-ln}e^{bq^l}
}





When we take the limit $S\rightarrow \infty$, we expect the $q$-deformed PB Fock space to restore to the SG Fock space. This suggests that the two-mode squeezed states constructed on the $q$-deformed Fock space should restore to the SG two-mode squeezed states. 
We will test whether it is true by looking at the entanglement entropy $E$ of $|r\rangle_q^S$ compared with $|r\rangle$ for increasing $S$. 


\subsection{The entanglement entropy of $|r\rangle_q^S$ compared with $|r\rangle^S$ }

\begin{figure}[H]
\begin{center}
\includegraphics[scale=0.5]{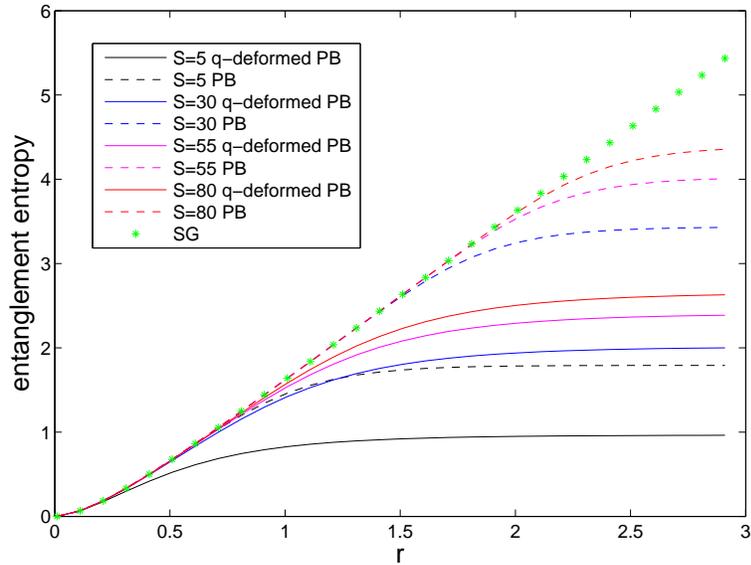}\caption{\label{fig:type2}
$E(|r\rangle_q^S)$ compared to $E(|r\rangle^S)$ and $E(|r\rangle)$}.
\end{center}
\end{figure}

\begin{figure}[H]
\begin{center}
	 \includegraphics[scale=0.5]{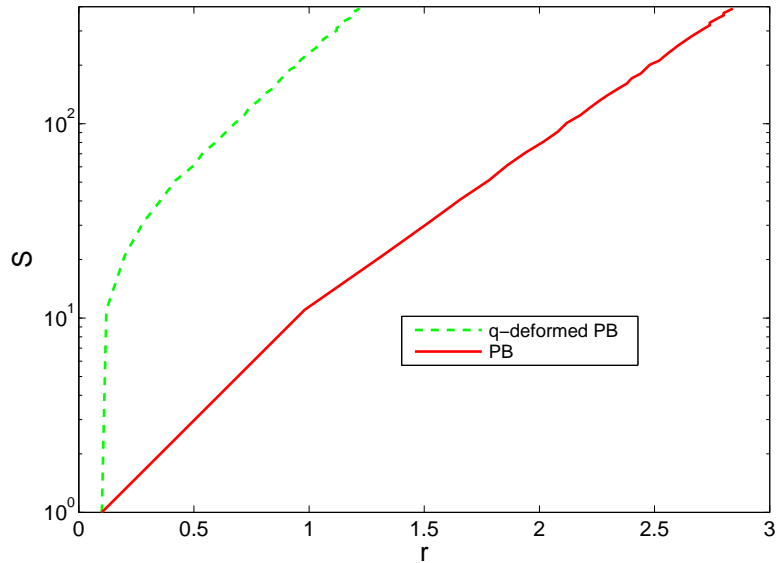}	
\caption{\label{fig:error} Value of $S$ required for less than $0.5\%$ error in the approximation of $E(|r\rangle)$ using $E(|r\rangle_q^S)$ and $E(|r\rangle^S)$. }
\end{center}
\end{figure}

Entanglement has been regarded as a useful resource for quantum information processing.  For qubits, entanglement is typically measured by considering the entropy of the reduced density matrix of a bipartite system. Analogous to Shannon entropy in information theory, 
the von Neumann entropy $S$ of a state with density operator $\rho$ is defined as \cite{Nielsen&Chuang}
\eq{
S(\rho)=-\tr(\rho \log \rho),
}
where the logarithms are taken to base two. 
If $\lambda_i$ are the eigenvalues of $\rho$ then the von Neumann entropy can be re-expressed
\eq{
S(\rho)=-\sum_{i} \lambda_i \log \lambda_i.
}
For continuous variable, the quantum teleportation of unknown coherent states has been realized experimentally by exploiting two-mode squeezed states as a resource.   Similar to a bipartite system of qubit, for the two-mode squeezed state, which is a pure bipartite state, the amount of entanglement is quantified by the von Neumann entropy of the reduced state of one mode.\\\\
The density operator of the reduced state of one mode of $|r\rangle$ is 
\eq{
\rho_1=\sum_{n=0}^{\infty}|c_{n}|^2|n\rangle \langle n| 
}
where $c_n=\frac{tanh^n(r)}{cosh(r)}$.\\
Therefore the entanglement of the two-mode squeezed state is \cite{Hiroshima01}
\eqa{\nonumber
E(|r\rangle)&=&S(\rho_1)\\\nonumber
&=&-\sum_{n=0}^{S}|c_{n}|^{2}\log|c_{n}|^{2}\\
&=&\cosh^{2}(r)\log[\cosh^{2}(r)]-\sinh^{2}(r)\log[\sinh^{2}(r)].
}
In the PB Fock space, for the maximally entangled bipartite pure state, 
\eq{
\rho_{max}=\sum_{n=0}^{S}\frac{1}{S+1}|n\rangle^S {}^S\langle n| \\
}
is the density operator of its reduced state of one mode. \\
The entanglement is maximum for the maximally entangled state and its value is
\eq{
E_{max}(S)=\log(S+1).
}
In \fref{fig:type2}, we compare the behaviour of $E(|r\rangle_q^S)$ with $E(|r\rangle^S)$ and $E(|r\rangle)$ for increasing $S$. 
\Fref{fig:type2} shows that $E(|r\rangle^S)$ approximates $E(|r\rangle)$ better than $E(|r\rangle_q^S)$ for the same $S$.
With a fixed $S$,  both $E(|r\rangle^S)$ and $E(|r\rangle_q^S)$ increases with $r$ when $r$ is small and is bounded for large $r$.  With increasing $S$, both $E(|r\rangle^S_q)$ and  $E(|r\rangle^S)$ are able to approximate $E(|r\rangle)$ for larger range of $r$. We may infer that $|r\rangle_q^S$ can approximate $|r\rangle$ for any $r$ if $S$ is large enough. To test numerically, we plot the value of $S$ required for  $E(|r\rangle_q^S$ and $E(|r\rangle)$ to be an approximation to $E(|r\rangle)$ within $0.5\%$ error in \fref{fig:error}.

\Fref{fig:error} shows that with increasing $r$, the value of $S$ required for $E(|r\rangle_q^S)$ and $E(|r\rangle^S)$ to approximate $E(|r\rangle)$ increases more rapidly for the former (deformed states) than the latter.

In \fref{fig:type2} we have already seen that for a given $S$, $E(|r\rangle_q^S)$ and $E(|r\rangle^S)$ increase and saturate when $r$ increases. To compare the saturation values with the maximum entanglement $E_{max}(S)=\log(S+1)$, $\frac{E(|r\rangle_q^S)}{E_{max}(S)}$ and $\frac{E(|r\rangle^S)}{E_{max}(S)}$ are plotted against $S$ with a very large  $r$ ($r$=20).
\begin{figure}[H]
\begin{center}
\includegraphics[scale=0.75]{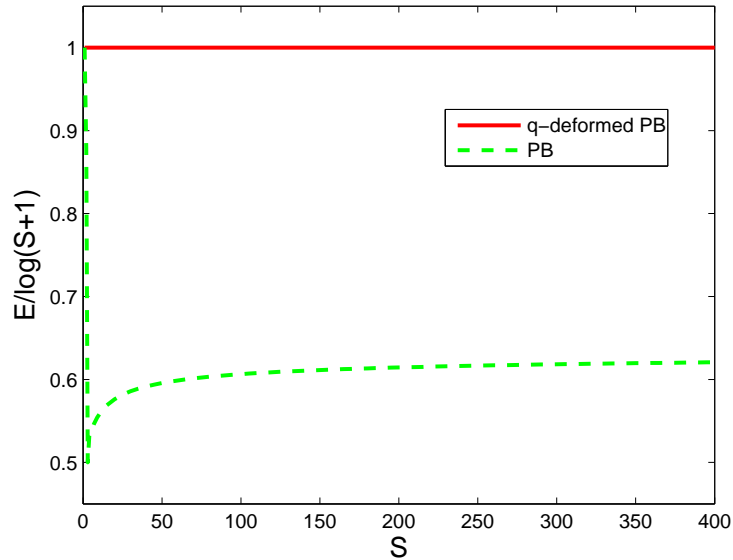}\caption{\label{fig:ratio} $\frac{E(|r\rangle_q^S)}{\log(S+1)}$ and $\frac{E(|r\rangle^S)}{\log(S+1)}$ vs. $S$, $r=20$. }
\end{center}
\end{figure}
\Fref{fig:ratio} shows that for $r=20$, $|r\rangle^S$ is maximally entangled for any $S$ in the range plotted on graph while $E(|r\rangle_q^S)$ increases with $S$ but saturates around $0.62\log(S+1)$. Therefore $|r\rangle_q^S$ never reaches the maximally entangled state for finite $S$ no mater how large $S$ is. 
Since $|r\rangle$ is a maximally entangled state when $r\rightarrow\infty$, we can conclude that $|r\rangle_q^S$ can never approach $\displaystyle{\lim_{r\to \infty}}|r\rangle$ although it can approximate $|r\rangle$ for any $r$ given that $S$ is large enough.

\section{Conclusion}

We have studied the coherent state and two-mode squeezed state in SG, PB , $q$-deformed, and $q$-deformed PB formalism. In particular, we have shown that the $q$-deformed states (coherent and squeezed) display different characteristics for finite truncation $S$ compared to the Pegg-Barnett states.  We look at the entanglement properties of two-mode squeezed states for both the $q$-deformed PB and PB squeezed states with the squeezing parameter $r$. For small $r$, there is agreement between the entanglement entropy of PB and SG formalism. Moreover, we see that if the truncation $S$ is sufficiently large, one could in principle approximate the SG state with the PB (deformed and undeformed) states for arbitrary $r$.  However, the value of $S$ required increases more rapidly with the $q$-deformed PB states than the PB states as a function of $r$.

\ack
We wish to acknowledge the funding support for this project from Nanyang Technological University under the Undergraduate Research Experience on CAmpus (URECA) programme.

\section*{References}
\bibliographystyle{unsrt}
\bibliography{myref}

\end{document}